# Micromagnetic simulations of persistent oscillatory modes excited by spin-polarized current in nanoscale exchange-biased spin valves

G. Siracusano[1], G. Finocchio[1], I. N. Krivorotov[2], L. Torres[3], G. Consolo[1], B. Azzerboni[1]

[1] Department of Fisica della Materia e Ingegneria Elettronica. University of Messina. Salita Sperone 31, 98166, Messina Italy.
[2] Department of Physics and Astronomy, University of California, Irvine, 92697-4575 CA, USA.
[3] Department of Física Aplicada, Universidad de Salamanca, Plaza de la Merced, Salamanca 37008, Spain.

**Abstract** – We perform 3D micromagnetic simulations of current-driven magnetization dynamics in nanoscale exchange biased spin-valves that take account of (i) back action of spin-transfer torque on the pinned layer, (ii) non-linear damping and (iii) random thermal torques. Our simulations demonstrate that all these factors significantly impact the current-driven dynamics and lead to a better agreement between theoretical predictions and experimental results. In particular, we observe that, at a non-zero temperature and a sub-critical current, the magnetization dynamics exhibits non-stationary behaviour in which two independent persistent oscillatory modes are excited which compete for the angular momentum supplied by spin-polarized current. Our results show that this multi-mode behaviour can be induced by combined action of thermal and spin transfer torques.

A spin-polarized current can apply a torque to nanomagnets by the direct transfer of spin angular momentum.[1,2] Experiments have shown that dc spin currents can drive magnetization reversal between static configurations or can excite the magnetization into steady-state dynamical modes.[3-6] For applied field large enough to saturate the magnetization of the nanomagnet along a given direction, a wide range of currents excites modes of large-amplitude magnetization oscillation. For example, in 130 x 70 x 3 nm$^3$ elliptical nanomagnets, dynamical stability diagram (frequency spectra as a function of current and out of plane external field) show the existence of frequency jumps of the excited modes. These jumps can be attributed to the reversal of the free layer precession axis from nearly parallel to nearly anti-parallel to the applied field direction.[7]

In contrast, in exchange biased spin valves (Py (4nm)/Cu(8nm)/Py(4nm)/IrMn(8nm)) (Py=Ni$_{80}$Fe$_{20}$) of elliptical cross sectional area (130 nm x 60 nm) for an in-plane applied field and a range of currents, the spectra show two independent modes at similar frequencies (see Fig.2c in Ref[8] I=3.5–5 mA). Measurements of these dynamics in frequency and time domains reveal that the magnetization dynamics are characterized by nanosecond scale switching between those modes. This rapid switching behaviour is also supported by the presence of a low power frequency peak below 1GHz seen in the experimental data.[9] Results of comparisons between experimental data and simulations show that while some experimental features can be reproduced, important quantitative differences between theory and experiment remain.[8,10-12] The origin of the quantitative differences between theory and experiment may lie in the simplifying assumptions used in most micromagnetic simulations of current-driven dynamics reported up to date. These simplifying assumptions include: one-dimensional character of spin and charge currents in the nanopillar (uniform throughout the spin valve structure), immobile fixed layer of the spin valve and decoupling of electrical transport and magnetization dynamical parts of the problem.

In this paper we report results of advanced micromagnetic simulations of large amplitude current-driven magnetization dynamics in exchange biased samples which go beyond the usual oversimplified approach to the problem. In particular, we study the effects of: (i) the back action of the torque to the pinned layer; (ii) the effect of non-linear damping and (iii) a spin torque with a stochastic component (we include thermal fluctuation in the pinned layer) on magnetization dynamics in nanoscale spin valves. Finally, by means of the micromagnetic spectral mapping technique (MSMT)[13], we describe the non-stationary behaviour of magnetization dynamics.

Our approach is to solve numerically the Landau-Lifshitz-Gilbert-Slonczewski equation considering the entire spin valve device[1,14,15]:

$$\begin{cases} \dfrac{d\mathbf{m_f}}{d\tau} = -(\mathbf{m_f} \times \mathbf{h_{f\text{-}eff}}) + \alpha_G \mathbf{m_f} \times \dfrac{d\mathbf{m_f}}{d\tau} - \mathbf{T}(\mathbf{m_p}, \mathbf{m_f}) \\ \dfrac{d\mathbf{m_p}}{d\tau} = -(\mathbf{m_p} \times \mathbf{h_{p\text{-}eff}}) + \alpha_G \mathbf{m_p} \times \dfrac{d\mathbf{m_p}}{d\tau} - \mathbf{T}(\mathbf{m_p}, \mathbf{m_f}) \end{cases} \quad ; \quad \mathbf{m} = \begin{cases} \mathbf{m_f} \\ \mathbf{0} \\ \mathbf{m_p} \end{cases} \quad (1)$$

where $\mathbf{m}$ is the magnetization of the whole device, $\mathbf{m_f} = \mathbf{M_f}/M_S$ is the normalized magnetization of the free layer, $\mathbf{m_p} = \mathbf{M_p}/M_S$ is the normalized magnetization of the pinned layer, $\mathbf{m}=\mathbf{0}$ in the copper spacer, $M_S$ is the saturation magnetization. $d\tau = \gamma_0 M_S dt$ is the dimensionless time step, $\alpha_G$ is the Gilbert damping parameter. $\mathbf{h}_{f\text{-}eff}$ and $\mathbf{h}_{p\text{-}eff}$ are the dimensionless effective fields for the free and the pinned layer respectively. The effective fields include the standard micromagnetic contributions from exchange ($\mathbf{h}_{exch}$), anisotropy ($\mathbf{h}_{ani}$), external $\mathbf{h}_{ext}$, and demagnetizing fields $\mathbf{h}_M$ ($\mathbf{h}_M$ is computed considering the magnetization of both the pinned and free layer). Furthermore we include the Oersted field due to the current.[15] The spin transfer torque is implemented according to the following equation:

$$\mathbf{T}(\mathbf{m_p}, \mathbf{m_f}) = \dfrac{g\,|\mu_B|}{e\gamma_0\,M_s^2} \begin{cases} \dfrac{j}{L_F}\varepsilon(\mathbf{m_f},\mathbf{m_p})\;\mathbf{m_f}\times(\mathbf{m_f}\times\mathbf{m_p}) & \text{free layer} \\ 0 & \text{copper spacer} \\ \dfrac{-j}{L_P}\varepsilon(\mathbf{m_p},\mathbf{m_f})\;\mathbf{m_p}\times(\mathbf{m_p}\times\mathbf{m_f}) & \text{pinned layer} \end{cases} \quad (2)$$

where $g$ is the gyromagnetic splitting factor, $\gamma_0$ is the gyromagnetic ratio, $\mu_B$ is the Bohr magneton, $j$ is the current density assumed to be spatially uniform over the entire device, $L_f$ and $L_P$ are the thickness of the free and pinned layer respectively, $e$ is the electron charge, and $\varepsilon(\mathbf{m_f},\mathbf{m_p})$ is the polarization function which characterizes the angular dependence of the Slonczewski spin torque term computed for symmetric layer[1]:

$$\varepsilon(\mathbf{m_f},\mathbf{m_p}) = \varepsilon(\mathbf{m_p},\mathbf{m_f}) = 0.5 P \Lambda^2 / \left(1+\Lambda^2 + (1-\Lambda^2)\mathbf{m_p}\cdot\mathbf{m_f}\right) \quad (3)$$

where $\Lambda^2 = \chi + 1$, $\chi$ is the giant-magneto-resistance (GMR) asymmetry parameter, $P$ is the current spin-polarization factor. By convention, positive current polarity corresponds to electron flow from the free to the pinned layer of the spin valve.

We employ device parameters similar to those used in experiments on current-driven dynamics in exchange-biases spin valves.[5,8,9] We use $M_S = 650 \times 10^3$ A/m, $\chi = 1.5$ and $P = 0.38$, we employed the standard exchange strength for Py $A = 1.3 \times 10^{-11}$ J/m. We use for the free layer a Gilbert damping parameter $\alpha_G = 0.025$ that was measured by a time-domain technique described in [5], for the pinned layer $\alpha_G$ is 0.2 because of the coupling with the antiferromagnet that gives a giant enhancement of the Gilbert damping in that ferromagnet.[16] We apply an in-plane external field of 68mT along -45 degree with respect to the easy axis of the ellipse, and we modeled the presence of the antiferromagnet as an additive field (75mT) applied to the pinned layer along 45 degree with respect to the easy axis of the ellipse. To include thermal effects in our simulations we add a random thermal field $\mathbf{h}_{th}$ to the $\mathbf{h}_{eff}$ for each micromagnetic cell (pinned and free layer).[17,18] For our simulations, we use a computational time step of 28 fs, and a spatial discretization cell of $(5 \times 5 \times 4\ nm^3)$; tests with a cell size of $(2.5 \times 2.5 \times 4.0\ nm^3)$ gave very similar results. In the z direction we have 4 discretization layer, the top and the bottom are related to the magnetization of the free and pinned layer respectively.

We performed a systematic numerical study of large-amplitude magnetization dynamics driven by spin-polarized current in these spin valve devices. Our simulations demonstrate that the non-stationary mode hopping behavior as described in [9] is excited by a combination of both random thermal fluctuations and deterministic spin-transfer-torque. Removing the thermal field the magnetization precession is first excited at 4.4 mA - a value larger than the experimental one (I=2.8mA).[8,9] At currents above the value 4.4mA, the magnetization precession is mainly characterized by a single peak in the frequency domain for a wide range of currents (we simulated up to 12.2 mA). Fig. 1(a)-(c) show the simulation results for a current of 6.15mA: (a) the temporal evolution of the average normalized magnetization components (blue <x>, red <y>, black <z>), (b) the frequency spectra of the x component of the magnetization of the free layer computed by means of the MSMT as can be noted the frequency spectrum is very close to the experimental one (see Fig.2b in Ref[8]). Fig. 1(c) shows the temporal evolution of the average normalized magnetization of the pinned layer (blue <x>, red <y>, black <z>) for the same process. Our important result is the presence of oscillations in the pinned layer due to the magnetostatic coupling between the layers and the back spin-torque effect. The frequency of the excited mode in the pinned layer is the same as that of the free layer. A complete comparison of the results of micromagnetic simulations (I=4.4-12.2mA) of the entire spin valve device (ED) with the results obtained considering the immobile pinned layer (PL-F) show the frequency of the mode excited in the ED configuration is larger than the frequency of the modes excited with the PL-F (e.g. I=6.15mA, $f_{ED}$=4.84GHz, $f_{PL-F}$=4.65GHz). The precession frequency in the ED configuration increases because the time-average effective field along the oscillation axis acting on the free layer magnetization is larger in this configuration. This is due to a reduction of the time-average dipolar field acting on the free layer from the moving fixed layer in the ED configuration (see Fig.1 of Ref. 9 for the details of the configuration).

Fig. 1 (d) shows the ED simulations including the thermal fluctuations computed for I=6.15mA and T=25K, our results show that the thermal fluctuations give rise to a more noise output signal which however does not substantially alter the large-amplitude magnetization dynamics (in both time and frequency domains).

A detailed comparison between experimental data and micromagnetic simulations for exchange biased spin valve structures was previously carried out in Ref. [8]. Although most features of the experimental data are well reproduced by simulations in Ref. [8], significant quantitative discrepancies for the amplitude of the excited modes (in terms of integrated root mean square amplitude spectral density) as a function of current are observed.[8] We performed generalized simulations including the non-linear damping $\alpha(\zeta) = \alpha_G \left[1 + q_1 \zeta^2\right]$, where $\zeta$ is the dimensionless variable proportional to the amplitude of magnetization precession, $q_1$ is a dimensionless phenomenological nonlinear damping parameter.[19] The results are summarized in Fig. 1(e)-(h). Figs. (e) and (g) show the temporal evolution of the three components of the average normalized magnetization (blue <x>, red <y>, black <z>) for $q_1=1$ and $q_1=3$. As can be noted from the spectra displayed in Fig.1(f) and Fig.1(h), as $q_1$ increases: (i) the frequency of the spectral peak increases and (ii) the integrated root mean square amplitude of the spectral density decreases. We conclude that the non-linear damping parameter gives rise to a trend in the amplitude of oscillations to slowly increase as a function of current in agreement with the experimental data.[8]

ED simulations that include thermal fluctuations for currents up to 4.4 mA give spectra with two peaks at nearby frequencies. Fig. 2(a) and (b) show the temporal evolution of the x-component of the average normalized magnetization (10-30ns) and the corresponding spectrum computed with the MSMT for a current I=3.1mA.

The dynamics is characterized by two primary excited modes: $P_1$ at 6.3GHz (almost uniform mode) and $P_2$ at 7.4GHz (edge mode). Performing the MSMT in different time intervals, we observe the magnetization dynamics are governed by the $P_1$ mode in the range 23.5-25.5ns and by the $P_2$ mode in the range 18-20ns. Those numerical results also show the existence of jumps between different dynamical modes (mode hopping) in the nanosecond regime in agreement with experiments.[9]

In summary, we performed micromagnetic simulations of large-amplitude current-induced magnetization dynamics in nanoscale spin valve devices. Our simulations take into account current-induced motion of not only the free but also the pinned ferromagnetic layer of the spin valve. We find that the motion of the pinned layer has significant impact on both the frequency and the amplitude of the microwave signal emitted by device. We have also determined that inclusion of the non-linear damping in the simulations decreases the amplitude of magnetization oscillations. Our simulations show that at a sub-critical current the magnetization oscillations show thermally-induced mode hopping behaviour at the nanosecond time scale.

FIG.1 : (a) temporal evolution of the $<\mathbf{m_f}>$ (blue $<x>$, red $<y>$, black $<z>$) of the average magnetization components of the free layer for I=6.15mA and T = 0 K; (b) corresponding spectrum computed by means of the MSMT of the $\mathbf{m_{f-x}}$; (c) temporal evolution of the $<\mathbf{m_p}>$ of the pinned layer for the the magnetization dynamics of Fig.1(a); (d) temporal evolution of the $<\mathbf{m_f}>$ of the free layer for I=6.15mA and T=25K; (e) temporal evolution of the $<\mathbf{m_f}>$ of the free layer for I=6.15mA and $q_1$=1; (f) corresponding MSMT spectrum of the $\mathbf{m_{f-x}}$; (g) temporal evolution of the $<\mathbf{m_f}>$ of the free layer for I=6.15mA and $q_1$=3; (h) corresponding MSMT spectrum of the $\mathbf{m_{f-x}}$;

FIG. 2: (a) temporal evolution of the $<\mathbf{m_{f-x}}>$ of the free layer for I=3.1mA and (b) relative spectrum computed by means of the MSMT; spectrum computed for the same process of Fig. 2(a) in short time intervals (c) 23.5-25.5ns; (d) 18-20ns.

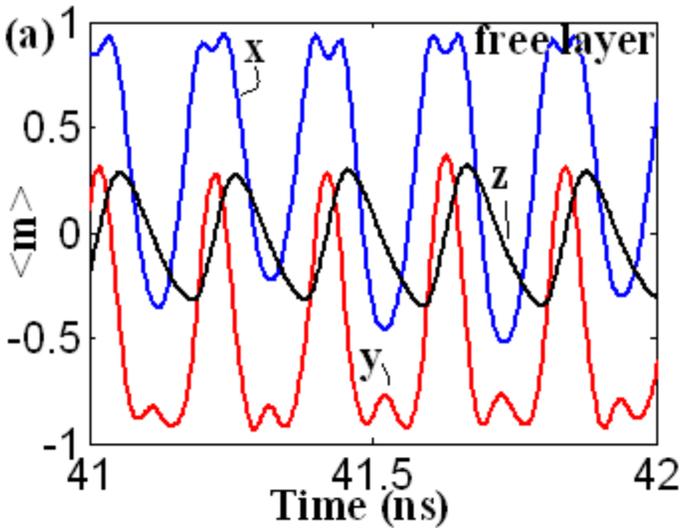
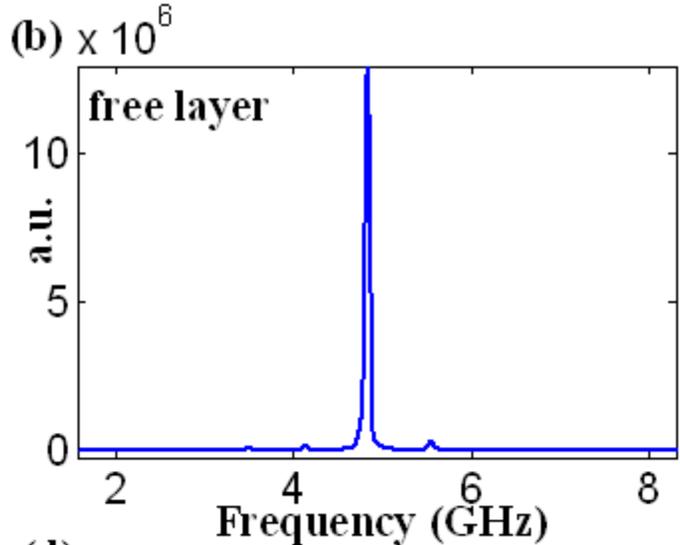
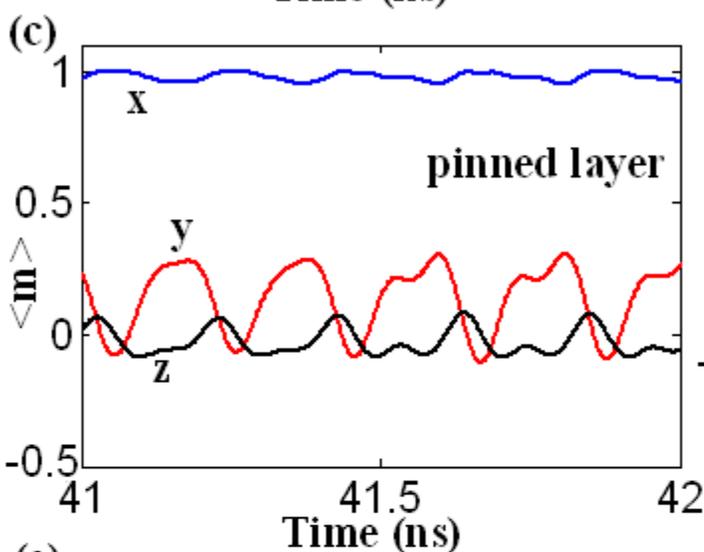
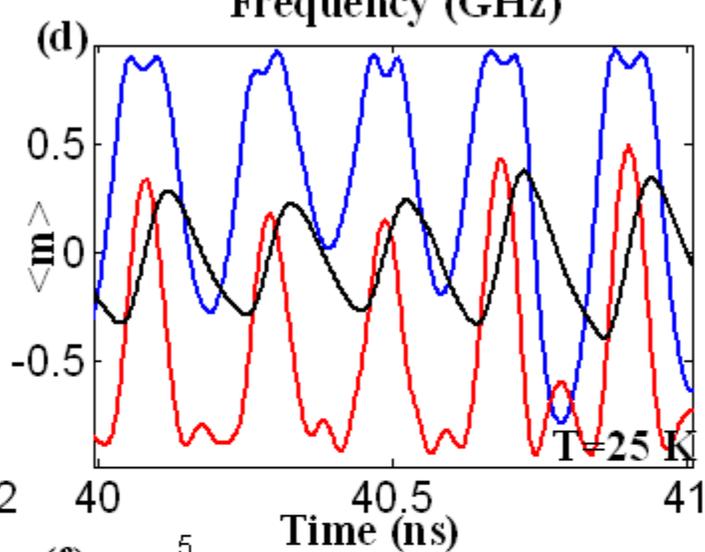
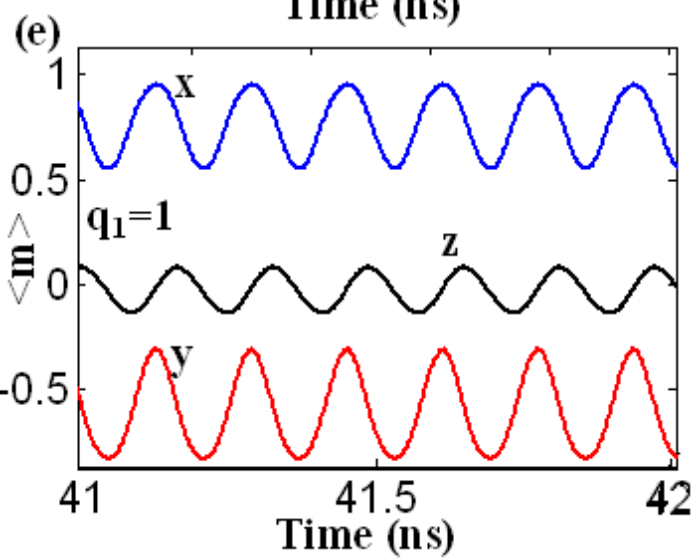
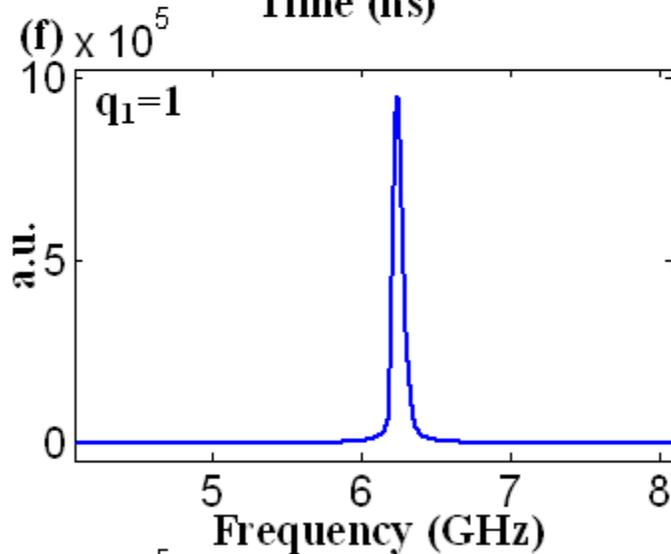
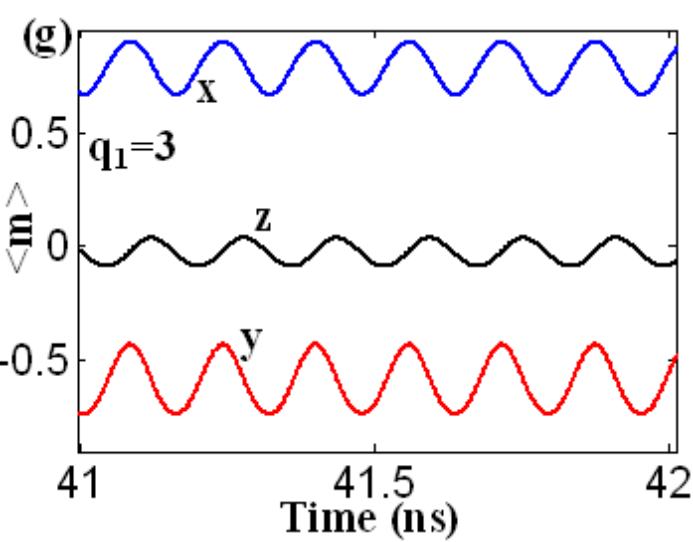
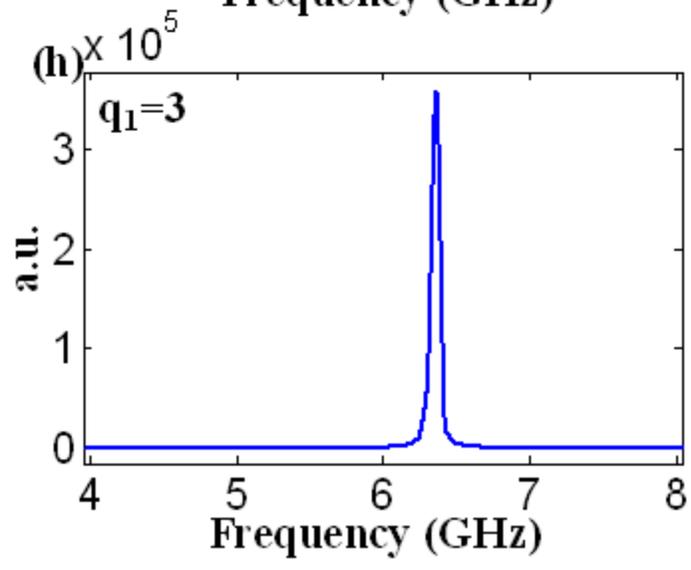

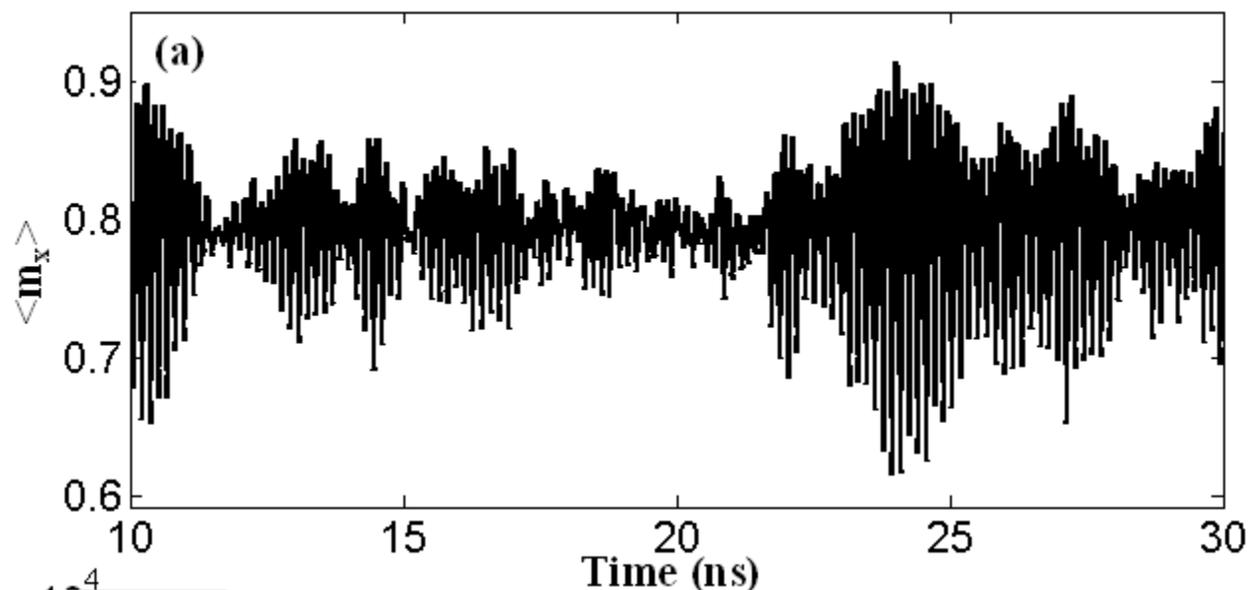